\magnification\magstep1 
\parskip=0.5 true cm
                           
  \def\sa{\vskip 0.30 true cm}
  \def\sb{\vskip 0.60 true cm}
  \def\sc{\vskip 0.15 true cm}
% \nopagenumbers 
       
% \pageno = 0 
% \voffset = - 5   true mm
  \hoffset =   1.5 true cm 
  \hsize = 15.5   true cm
% \vsize = 25.2   true cm

%  \font\msym=msym10

% \font\msim=msym10

\baselineskip = 0.6 true cm

\rightline{\bf LYCEN 9148}
\rightline{October 1991}

\sa
\sb
\sa

\noindent {\bf TWO-PHOTON SPECTROSCOPY OF TRANSITION-METAL}

\noindent {\bf IONS IN CUBICAL SYMMETRY$^{*+}$}

\sa
\sb

\noindent M.~Daoud and M.~Kibler

\sa

\noindent Institut de Physique Nucl\'eaire, 
IN2P3-CNRS et Universit\'e Lyon-1, 69622 Villeurbanne Cedex 
(France) 

  \sa
  \sa
% \sb
% \sb

\baselineskip = 0.65 true cm

\sa

\noindent {\bf Summary} 

\sc

Symmetry adaptation techniques are applied to the determination 
of intensities of intra-configurational two-photon transitions 
for transition-metal ions in cubical symmetry. This leads to a 
simple model giving the polarization dependence of intensities 
of two-photon (electric dipolar) transitions between Stark levels 
of the configuration $3d^N$ ($N$ $=$ even or odd). 

\sa
\sb
\sa
\sb
\sa
\sb
\baselineskip = 0.5 true cm

\noindent $^*$ Communication at the ``5\`eme Rencontre 
Marocaine sur la Chimie de l'Etat Solide'', organized by the 
Facult\'e des Sciences I and the Facult\'e des Sciences II of 
the Universit\'e Hassan II and the Laboratoire Public d'Essais 
et d'Etudes (L.P.E.E), Casablanca, Morocco, 30 October~-~1 November 1991.

\noindent $^+$ 
Published in {\bf Journal of Alloys and Compounds 188 (1992) 255}. 
All correspondence concerning this paper 
should be addressed to~: M. Kibler, Institut de Physique Nucl\'eaire, 
Universit\'e Lyon-1, 43 Bd du 11 Novembre 1918, 69622 Villeurbanne 
Cedex (France). (Telephone~: 72 44 82 35. Fax~: 72 44 80 04. 
Electronic mail~: Kibler@frcpn11.) 

\vfill\eject
\baselineskip = 0.75 true cm

\noindent {\bf 1. Introduction}

With the availability of tunable dye lasers, two-photon 
spectroscopy of transition ions in molecular or solid-state 
environments (generic symmetry $G$) has been the object of 
numerous developments both from an experimental and theoretical 
viewpoint [1-23]. Indeed, two-photon spectroscopy turns out to 
be a useful complement to one-photon spectroscopy because it 
allows levels to be reached which cannot be seen in one-photon 
spectroscopy [1-13]. There are now many two-photon absorption spectra 
published for rare-earth ions (generic configuration $4f^N$ for 
the lanthanides or $5f^N$ for the actinides) in various 
surroundings. In this respect, to name but a few recently 
published studies, let us mention 
Gd$^{3+}$ ($N=7$) in Gd(OH)$_3$ ($G = D_{3h}$) [14,15], 
                  in GdCl$_3$   ($G = C_{3h}$) [14] and 
                  in the cubic elpasolite 
               Cs$_2$NaGdCl$_6$ ($G = O_{h}$) [16]~; 
Sm$^{2+}$ ($N=6$) in BaClF    ($G = C_{4v}$) [17,18] and 
                  in SrClF    ($G = C_{4v}$) [19]~; and, finally, 
Eu$^{3+}$ ($N=6$) in LuPO$_4$ ($G = D_{2d}$) [19]. 
Transition-metal ions of the iron series (generic configuration 
$3d^N$) in crystals have also been the object of recent 
investigations. For example, the case of Ni$^{2+}$ ($N=8$) in MgO 
($G = O)$ has received a great deal of attention in the last 
three years [20,21,22]. Furthermore, there are also some data 
about Co$^{2+}$ ($N = 7$) in KZnF$_3$ ($G = O$) [23]. 

It is the aim of the present paper to report on a simple model 
for describing $3d^N \to 3d^N$ intra-configurational two-photon 
transitions for a transition-metal ion 
in an environment of cubical (octahedral) symmetry ($G = O$ or 
$O_h$). The main ingredients of the model (symmetry adapted 
wave-functions and second- plus third-order mechanisms) are 
given in section 2 with necessary formulas for application 
listed in section 3.

\noindent {\bf 2. Theory}

Let us consider a (parity-allowed) two-photon transition 
between an initial state $i$, of symmetry $\Gamma$, and a final 
state $f$, of symmetry $\Gamma'$, of the configuration $3d^N$~; 
the labels $\Gamma$ and $\Gamma'$ stand for irreducible 
representations of the group $O$ or its double group $O^*$ 
according to whether the number $N$ of $3d$ electrons is even or 
odd, respectively. The corresponding state vectors are denoted 
$\vert 3 d^N i \Gamma  \gamma )$ and 
$\vert 3 d^N f \Gamma' \gamma')$ where $\gamma$ and $\gamma'$ 
distinguish the various partners for $\Gamma$ and $\Gamma'$. 
These vectors can be expressed either in a 
strong-field basis of type 
$\vert t_2^{N-M} (S_1 \Gamma_1) 
       e^M       (S_2 \Gamma_2) S_T \Gamma_T \beta \Gamma \gamma )$ 
or in a weak-field basis of type 
$\vert 3d^N \alpha S L J a \Gamma \gamma )$. We shall adopt 
here a weak-field approach. (The weak- and strong-field 
approaches are equivalent inso far as we use 
the same time-independent Hamiltonian 
$H_{i+e}$ for the ion in its environment [24] in both approaches.) 

We have chosen to calculate the transition matrix element 
$M_{i(\Gamma  \gamma ) \to 
    f(\Gamma' \gamma')}$ within the following approximations~: 
(i) We use single-mode excitations 
(energy      $\hbar  \omega_{\lambda}$, 
 wave-vector ${\vec      k}_{\lambda}$, 
polarization $\vec {\cal E}_{\lambda}$) of the radiation 
field and we suppose the two photons ($\lambda = 1,2$) to be 
identical. (In fact, most of the experiments achieved up to now 
use a single laser beam.) (ii) We use a time-dependent Hamiltonian
of the type $H_{i+e} + H_{rf} + H_{int}$ for describing the 
system formed by the ion in its environment ($H_{i+e}$) and the 
radiation field ($H_{rf}$) which interact through $H_{int}$ and 
we treat $H_{int}$ in the framework of the electric dipolar 
approximation. (iii) We use a quasi-closure approximation to 
deal with the G\"oppert-Mayer formula for two-photon processes. 

As a result, the transition matrix element can be calculated to 
be [1,5,6]
$$
M_{i (\Gamma \gamma) \to f (\Gamma' \gamma')} \; = \; 
(3 d^N f \Gamma' \gamma' \vert H_{eff} \vert 3 d^N i \Gamma \gamma) 
$$
where the effective operator $H_{eff}$ reads [10] 
$$
H_{eff} \; = \; \sum_{k = 0,2} \; \sum_{k_Sk_L} 
  \; C \left[ \left( k_S k_L \right) k \right] 
  \; \left( \left\{ {\cal E} \, {\cal E} \right\} ^{(k)} \, . \, 
  {\bf W}^{(k_Sk_L)k} \right) 
$$
The term $\left\{ {\cal E} \, {\cal E} \right\} ^{(k)}$ is 
the tensor product of rank $k$ of the polarization unit 
vectors ${\cal E}$ for the two photons. The 
dependence on the ion appears in the electronic double tensor 
${\bf W}^{(k_Sk_L)k}$ of spin rank $k_S$, orbital rank $k_L$ 
and total rank $k$. Furthermore, the 
$C \left[ \left( k_S k_L \right) k \right]$ parameters are 
expansion coefficients which may be calculated from 
first principles. The contributions $(k_S = 0, k_L = 2, k = 2)$ 
and $(k_S \ne 0, k_L, k)$ correspond to the standard 
second-order mechanisms [1-4] and to the so-called 
third-order mechanisms (which may take into account 
ligand, crystal-field and spin-orbit effects) [5-9], respectively. 

The intensity of the $i(\Gamma) \to f(\Gamma')$ two-photon transition, 
viz., 
$$
S_{  \Gamma  \to   \Gamma' } \; = \; \sum_{\gamma \gamma'} \; 
\left\vert M_{i(\Gamma \gamma) \to f(\Gamma' \gamma')} \right\vert ^2 
$$
can be calculated by using the symmetry adaptation techniques developed 
in refs.~24 and 25. We thus obtain 
$$
S_{\Gamma \to \Gamma'} \; = \; \sum_{k = 0,2} \; 
                            \sum_{\ell = 0,2} \; \sum_{\Gamma''} \;
I [k \ell \Gamma'' ; \Gamma \Gamma'] \; \sum_{\gamma''} \; \left\{ 
{\cal E} \, {\cal E} \right\} ^{(k)}    _{ \Gamma'' \gamma''} \; 
\left( \left\{ 
{\cal E} \, {\cal E} \right\} ^{(\ell)} _{ \Gamma'' \gamma''}\right)^* 
$$
where the $I$ parameters, which bear the dependence on the ion 
in its environment, have been derived 
in the weak-field coupling scheme [11-13,22]. These intensity 
parameters depend on the wave-functions used for the initial 
and final states, on the atomic parameters 
$C \left[ \left( k_S k_L \right) k \right]$, on atomic 
reduced matrix elements and on isoscalar factors for the 
chains of groups $SO(3) \supset O$ (for $N$ even) 
or $SU(2) \supset O^*$ (for $N$ odd). The number of independent 
parameters $I$ is controlled by a set of properties and rules 
[11-13]. For the purpose of this paper, it is sufficient to 
note that the sum over $\Gamma''$ is limited by the selection rule : 
$\Gamma''$ is of the type $A_1$, $E$ or $T_2$ and 
must be contained in the representation 
$\Gamma'^* \otimes \Gamma$ of the octahedral group $O$. In 
addition, the polarization dependence is completely contained 
in the factors of type 
$ \left\{ {\cal E} \, {\cal E} \right\} $.

\noindent {\bf 3. Application}

By applying the above-mentioned selection rule, we can rewrite 
$S_{\Gamma \rightarrow \Gamma '}$ as
$$
S_{\Gamma \rightarrow \Gamma '} \; = \;
{1 \over 3} \; I[00A_1 ; \Gamma \Gamma '] \; {\varpi_1} \; + \; 
{1 \over 6} \; I[22E   ; \Gamma \Gamma '] \; {\varpi_2} \; + \;
{1 \over 4} \; I[22T_2 ; \Gamma \Gamma '] \; {\varpi_3} 
$$
where the functions ${\varpi_i}$ ($i=1,2,3$) can be 
readily derived by means of Wigner-Racah calculus for the chain of 
groups $SO(3) \supset O$ [24,25]. As a matter of fact, we obtain
$$
\eqalign{
{\varpi_1} \; &= \; 3 \; 
\left | \{ {\cal {E}} {\cal {E}} \}^{(0)}_{A_1} \right |^2 
\; = \; 1 
\qquad {\hbox {or}} \qquad 0 \cr
{\varpi_2} \; &= \; 6 \; \sum_{\gamma''} \;
\left | \{ {\cal {E}} {\cal {E}} \}^{(2)}_{E \gamma ''} \right |^2
\; = \; 
(3 \cos^2 \theta - 1)^2 + 3 \sin^4 \theta \; \cos^2 2 \varphi 
\qquad {\hbox {or}} \qquad 3 \cr 
{\varpi_3} \; &= \; 4 \; \sum_{\gamma''} \;
\left | \{ {\cal {E}} {\cal {E}} \}^{(2)}_{T_2 \gamma ''} \right |^2
\; = \; 2 \; (\sin^4 \theta \; \sin^2 2 \varphi + \sin^2 2 \theta) 
\qquad {\hbox {or}} \qquad 2 
} 
$$
according to whether as the polarization is linear or 
circular. For linear polarization, $(\theta, \varphi)$ 
are the polar angles of the polarization vector 
${\vec {\cal E}}$ 
with respect to the crystallographic axis and, for circular 
polarization, the wave-vector $\vec k$ is parallel to the 
crystallographic axis. (Of course, the angular functions $\varpi_i$ 
($i=1,2,3$) do not depend on the labels $\gamma''$, i.e., on 
the chain $SO(3) \supset G$$=$$O \supset G' \supset G''$ 
used in practical computations.) 

We give below the intensities $S_{\Gamma \to \Gamma'}$ 
for $N$ even ($\Gamma$ and $\Gamma'$ belong to $O    $) and 
for $N$ odd  ($\Gamma$ and $\Gamma'$ belong to $O^{*}$). To 
pass from $S_{\Gamma  \to \Gamma'}$ to 
          $S_{\Gamma' \to \Gamma }$, it is sufficient to change 
$\Gamma \Gamma'$ into $\Gamma' \Gamma$ in the intensity 
parameters $I$. For 
    $N$ even, the results are the following. 
$$S_{A_1 \rightarrow A_1} = {1\over 3} \; I [00 A_1 ; A_1A_1] \; 
{\varpi_1}, \qquad S_{A_1 \rightarrow A_2} = 0,$$
$$S_{A_1 \rightarrow E} = {1\over 6} \; I [22 E ; A_1E] \; 
{\varpi_2}, \qquad S_{A_1 \rightarrow T_1} = 0,$$
$$S_{A_1 \rightarrow T_2} = {1 \over 4} \; I [22 T_2 ; A_1T_2] \; 
{\varpi_3}, \qquad 
 S_{A_2 \rightarrow A_2} = {1\over 3} \; I [00 A_1 ; A_2A_2] \; 
{\varpi_1},$$
$$S_{A_2 \rightarrow E} = {1\over 6} \; I [22 E ; A_2E] \; 
{\varpi_2}, \qquad 
 S_{A_2 \rightarrow T_1} = {1 \over 4} \; I [22 T_2 ; A_2T_1] \; 
{\varpi}_3,$$
$$S_{A_2 \rightarrow T_2} = 0, \qquad 
 S_{E \rightarrow E} = {1\over 3} \; I [00 A_1 ; EE] \; {\varpi}_1 +
{1\over 6} \; I [22 E ; EE] \; {\varpi}_2,$$
$$S_{E \rightarrow T_1} = {1 \over 4} \; I [22 T_2 ; ET_1] \; 
{\varpi}_3, \qquad 
 S_{E \rightarrow T_2} = {1 \over 4} \; I [22 T_2 ; ET_2] \; 
{\varpi}_3,$$
$$S_{T_1 \rightarrow T_1} = {1\over 3} \; I [00 A_1 ; T_1T_1] \; {\varpi}_1 +
{1\over 6} \; I [22 E ; T_1T_1]   \; {\varpi}_2 +
{1 \over 4} \; I [22 T_2 ; T_1T_1] \; {\varpi}_3,$$
$$S_{T_1 \rightarrow T_2} = {1\over 6} \; I [22 E ; T_1T_2] \; {\varpi}_2 +
{1 \over 4} \; I [22 T_2 ; T_1T_2] \; {\varpi}_3,$$
$$S_{T_2 \rightarrow T_2} = {1\over 3} \; I [00 A_1 ; T_2T_2] \; {\varpi}_1 +
{1\over 6} \; I [22 E ; T_2T_2]   \; {\varpi}_2 +
{1 \over 4} \; I [22 T_2 ; T_2T_2] \; {\varpi}_3.$$
For $N$ odd, we have the following intensity formulas. 
$$S_{\Gamma_6 \rightarrow \Gamma_6} =
{1\over 3} \; I [00 A_1 ; \Gamma_6 \Gamma_6] \; {\varpi}_1, \qquad 
 S_{\Gamma_6 \rightarrow \Gamma_7} =
{1 \over 4} \; I [22T_2 ; \Gamma_6\Gamma_7] \; {\varpi}_3,$$
$$S_{\Gamma_6 \rightarrow \Gamma_8} =
{1 \over 6} \; I [22 E ; \Gamma_6\Gamma_8]   \; {\varpi}_2 +
{1 \over 4} \; I [22 T_2 ; \Gamma_6\Gamma_8] \; {\varpi}_3,$$
$$S_{\Gamma_7 \rightarrow \Gamma_7} =
{1\over 3} \; I [00 A_1 ; \Gamma_7\Gamma_7] \; {\varpi}_1, \quad 
 S_{\Gamma_7 \rightarrow \Gamma_8} =
{1 \over 6} \; I [22 E ; \Gamma_7\Gamma_8]   \; {\varpi}_2 +
{1 \over 4} \; I [22 T_2 ; \Gamma_7\Gamma_8] \; {\varpi}_3,$$
$$S_{\Gamma_8 \rightarrow \Gamma_8} =
{1 \over 3} \; I [00 A_1 ; \Gamma_8\Gamma_8] \; {\varpi}_1 +
{1 \over 6} \; I [22 E ; \Gamma_8\Gamma_8]   \; {\varpi}_2 +
{1 \over 4} \; I [22 T_2 ; \Gamma_8\Gamma_8] \; {\varpi}_3.$$

 The intensity formulas given above cover all the possible ground 
states encountered for transition-metal ions in cubical symmetry. 
We note that
$$
S_{A_1 \to A_1} = 
S_{A_2 \to A_2} = S_{\Gamma_6 \to \Gamma_6} = 
                  S_{\Gamma_7 \to \Gamma_7} = 0 
$$
when the scalar contribution (characterized by 
$I[00A_1 ; \Gamma \Gamma]$) to the third-order mechanisms is not 
taken into consideration. Therefore, the observation, if any, of the 
latter transitions would prove the relevance of third-order 
mechanisms. In particular, it would be interesting to test the 
importance of the third-order mechanisms in the case of an ion 
with configuration $3d^5$ (like Mn$^{2+}$). 

The expression of the intensity parameters $I$ 
has been described in section 2 in 
the weak-field coupling scheme. 
(They can be equally well expressed in the 
strong-field coupling scheme.) There 
are three ways to deal with the $I$ parameters. First, they may 
be considered as phenomenological parameters to be ajusted on 
experimental data. Second, they may be calculated from first 
principles. We then need to diagonalize-optimize the 
matrix of $H_{i+e}$ 
(as done, for instance, in ref.~26 for Eu$^{3+}$ in 
fifteen compounds of interest in solid-state chemistry) 
and to calculate isoscalar factors, reduced 
matrix elements and parameters characterizing 
second- and/or third-order mechanisms. Third, they may be 
handled in a mixed (semi-phenomenological) way, especially 
if we want to reduce the number of $I$ parameters.  

As an illustration, let us consider the case of Ni$^{2+}$ in 
MgO. The two-photon transitions from the initial state 
$i = {^3A_2}(T_2)$ with $\Gamma = T_2$ to the final states 
$f = {^3T_2}(E  )$ with $\Gamma'= E  $ and 
$f = {^3T_2}(T_1)$ with $\Gamma'= T_1$ have been recently 
observed for various linear polarizations [20-21]. The 
specialization to the configuration $3d^8$ of the model 
described here allows computation in an {\it ab initio} way of 
the intensity ratios $R_1$ and $R_2$ defined in ref.~21. The 
      theoretical values are $R_1 = 0.95$    and $R_2 = 1.04$, to be compared 
with the experimental values $R_1 = 1.5$$-$$3$ and $R_2 = 1.1 $ [22].

\vfill\eject
\noindent {\bf 4. Conclusion}

In this paper we have concentrated on the intensities of 
two-photon transitions for $3d^N$ ions in octahedral symmetry. 
The model discussed in section 2 is valid for any strength of 
the crystal-field interaction. Therefore, 
the results of sections 2 and 3 can be extended {\it mutatis 
mutandis} to any $nd^N$ configuration ($n = 4$ for the palladium 
series and $n = 5$ for the platinum series). They can be also 
applied to tetrahedral 
symmetry in view of the isomorphism of $O$ and $T_d$. Finally, the 
results given here concern one-color transitions. The extension 
to two-color transitions (using two different beams) is straightforward. 

A particular version of the 
model presented in this paper has been successfully applied 
to Ni$^{2+}$ in MgO~; the main results have been discussed at 
REMCES V and will be published elsewhere in greater detail [22]. The model 
will be applied to some other 
experimental data (e.g., Co$^{2+}$ in KZnF$_3$ [23]) in the thesis by 
one of us (M.~D.) and in forthcoming papers.  

\noindent {\bf Acknowledgments}

The authors thank G. Burdick, J.C. G\^acon, B. Jacquier, R. Moncorg\'e 
and J. Sztucki for interesting discussions. One of the authors 
(M.K.) is grateful to M. Reid and C. Campochiaro for 
correspondence. 

\vfill\eject
\baselineskip = 0.70 true cm

\noindent {\bf References}

\item{1} J.D. Axe, Jr., Phys. Rev., 136 (1964) A42.

\item{2} M. Inoue and Y. Toyozawa, J. Phys. Soc. Japan, 20 (1965) 363. 

\item{3} T.R. Bader and A. Gold, Phys. Rev., 171 (1968) 997.

\item{4} P.A. Apanasevich, R.I. Gintoft, V.S. Korolkov, A.G. 
Makhanek and G.A. Skrip\-ko, Phys. Status Solidi (b), 58 (1973) 
745~; A.G. Makhanek and G.A. Skrip\-ko, Phys. Status Solidi (a), 53 (1979) 
243~; A.G. Makhanek, V.S. Korolkov 
  and L.A. Yuguryan, Phys. Status Solidi (b), 149 (1988) 231.

\item{5} B.R. Judd and D.R. Pooler, J. Phys. C, 15 (1982) 591.

\item{6} M.C. Downer and A. Bivas, Phys. Rev. B, 28 (1983) 3677. 

\item{7} M.F. Reid and F.S. Richardson, Phys. Rev. B, 29 (1984) 2830. 

\item{8} J. Sztucki and W. Str\c ek, Chem.~Phys.~Lett., 125 (1986) 520.

\item{9} L. Smentek-Mielczarek and B.A. Hess, Jr., Phys. Rev., 
B 36 (1987) 1811. 

\item{10} M. Kibler and J.C. G\^acon, Croat. Chem. Acta, 62 (1989) 783. 

\item{11} M. Kibler, in W. Florek, T. Lulek and M. Mucha (eds.), 
{\it Symmetry and Structural Properties of Condensed Matter}, 
World Scientific, Singapore, 1991, p.~237. 

\item{12} M. Kibler and M. Daoud, {\it Proc.~V Workshop on 
Symmetry Methods in Physics, Obninsk, USSR, July 1991}, in the press. 

\item{13} M. Kibler, {\it Proc.~IInd International School on 
Excited States of Transition Elements, Karpacz, Poland, September 
1991}, World Scientific, Singapore, in preparation. 

\item{14} B. Jacquier, Y. Salem, C. Linar\`es, J.C. G\^acon, R. 
Mahiou and R.L. Cone, J. Lumin., 38 (1987) 258.

\item{15} B. Jacquier, J.C. G\^acon, Y. Salem, C. Linar\`es 
and R.L. Cone, J.~Phys.~: Condens. Matter, 1 (1989) 7385.

\item{16} M. Bouazaoui, B. Jacquier, L. Linar\`es, 
W. Str\c ek and R.L. Cone, J. Lumin., 48/49 (1991) 318. 

\item{17} J.C. G\^acon, J.F. Marcerou, M. Bouazaoui, B. Jacquier
and M. Kibler, Phys. Rev. B, 40 (1989) 2070.

\item{18} J.C. G\^acon, B. Jacquier, 
J.F. Marcerou, M. Bouazaoui and M. Kibler, J. Lumin., 45 (1990) 162.

\item{19} J.C. G\^acon, M. Bouazaoui, B. Jacquier, M. Kibler, L.A. 
Boatner and M.M. Abraham, Eur. J. Solid State Inorg. Chem., 28 (1991) 113. 

\item{20} R. Moncorg\'e and T. Benyattou, Phys. Rev. B, 37 (1988) 9186. 

\item{21} C. Campochiaro, D.S. McClure, P. Rabinowitz 
and S. Dougal, Phys. Rev. B, 43 (1991) 14. 

\item{22} J. Sztucki, M. Daoud and M. Kibler, Phys. Rev. B, 45 
(1992) 2023.

\item{23} R. Moncorg\'e, personal communication, 1991.

\item{24} M. Kibler and G. Grenet, {\it Studies in Crystal-Field 
Theory}, IPNL press, Lyon, 1986.

\item{25} M.~Kibler, C.~R.~Acad.~Sc.~(Paris) B, 268 (1969) 
1221~; M.R. Kibler and P.A.M. Guichon, Int.~J.~Quantum 
Chem., 10 (1976) 87~; M.R.~Kibler and G.~Grenet, Int.~J.~Quantum 
Chem., 11 (1977) 359~; M.R.~Kibler, Int. J. Quantum Chem., 23 (1983) 
115~; M.R.~Kibler, Croat.~Chem.~Acta 57 (1984) 1075.

\item{26} O.K. Moune, P. Caro, D. Garcia and M. Faucher, J. 
Less-Common Met., 163 (1990) 287. 

\bye